\documentclass[preprint,prl, preprintnumbers,superscriptaddress,amsmath,amssymb]{revtex4-1}

\usepackage[dvipdfmx]{graphicx}
\usepackage{colortbl}
\usepackage{mathrsfs}
\usepackage{lineno}
\usepackage{multirow}

\usepackage{bm} 
\usepackage{ulem}
\usepackage{amsmath}
\usepackage {autobreak}

\renewcommand{\Re}{\operatorname{Re}}
\renewcommand{\Im}{\operatorname{Im}}

\def\Vec#1{\mbox{\boldmath $#1$}}

\makeatletter
\def\btt#1{\texttt{\@backslashchar#1}}%
\DeclareRobustCommand\bblash{\btt{\@backslashchar}}%
\makeatother

\begin{document}


\title{Reconsidering the nonlinear emergent inductance: time-varying Joule heating and its impact on the AC electrical response}

\author{Soju Furuta}
\affiliation{Department of Physics, Tokyo Institute of Technology, Tokyo 152-8551, Japan}

\author{Wataru Koshibae}
\affiliation{RIKEN Center for Emergent Matter Science (CEMS), Wako 351-0198, Japan}

\author{Keisuke Matsuura}
\affiliation{Department of Physics, Tokyo Institute of Technology, Tokyo 152-8551, Japan}

\author{Nobuyuki Abe}
\affiliation{Department of Physics, College of Humanities and Sciences, Nihon University, Tokyo 156-8550, Japan}

\author{Fei Wang}
\affiliation{Department of Physics, Tsinghua University, Beijing 100084, People's Republic of China}

\author{Shuyun Zhou}
\affiliation{Department of Physics, Tsinghua University, Beijing 100084, People's Republic of China}

\author{Taka-hisa Arima}
\affiliation{RIKEN Center for Emergent Matter Science (CEMS), Wako 351-0198, Japan}
\affiliation{Department of Advanced Materials Science, University of Tokyo, Kashiwa 277-8561, Japan}

\author{Fumitaka Kagawa}
\email{kagawa@phys.titech.ac.jp}
\affiliation{Department of Physics, Tokyo Institute of Technology, Tokyo 152-8551, Japan}
\affiliation{RIKEN Center for Emergent Matter Science (CEMS), Wako 351-0198, Japan}

\date{\today}
\begin{abstract}
A nonlinearly enhanced electrical reactance, $\Im Z$, under a large AC  current has been measured to explore emergent inductors, which constitute a new class of inductors based on the spin-transfer torque effect. A nonlinear $\Im Z$ has been observed in conducting magnets that contain noncollinear magnetic textures and interpreted as the realization of an inductance due to current-induced spin dynamics. However, curious behavior has concomitantly been observed. For instance, the nonlinear $\Im Z$ always has a cutoff frequency of $10^0$--$10^4$ Hz, which is much lower than the resonance frequency of a ferromagnetic domain wall, $\sim$10$^7$ Hz. Furthermore, the magnitude of $\Im Z$ is much greater than that theoretically expected, and the temperature and magnetic field dependences are complicated. This behavior appears to be difficult to understand in terms of the current-induced spin dynamics, and therefore, the earlier interpretation of the nonlinear $\Im Z$ should be further verified.
Here, we theoretically and experimentally show that time-varying Joule heating and its impact on the AC electrical response can naturally explain these observations. In the experimental approach, we investigate the nonlinear AC electrical response of two conducting materials that exhibit no magnetic order, CuIr$_2$S$_4$ and 1$T$'-MoTe$_2$. Under time-varying Joule heating, a nonlinearly enhanced $\Im Z$ is observed in both systems, verifying the concept of the Joule-heating-induced AC electrical response. We reconsider the nonlinear emergent inductance reported thus far and discover that the Joule-heating-induced AC electrical response approximately reproduces the temperature and magnetic field dependences, cutoff frequency, and magnitude of $\Im Z$. Our study implies that the nonlinear $\Im Z$ previously observed in conducting magnets that contain noncollinear magnetic textures includes a considerable contribution of the Joule-heating-induced apparent AC impedance.

\end{abstract}

\maketitle


\section{I. INTRODUCTION}
The exchange of spin angular momentum between flowing conduction electrons and an underlying magnetic texture leads to the spin-transfer-torque (STT) effect on the magnetic texture \cite{STT1, STT2}. Nagaosa theoretically proposed a new class of inductors arising from the STT-induced elastic deformation of noncollinear magnetic textures \cite{NagaosaJJAP}, and these inductors are now referred to as emergent inductors. The time evolution of a magnetic texture creates an emergent electric field (EEF) \cite{Volovik, BarnesPRL2007}. Under an AC electric current below the threshold value, the magnetic texture remains in the pinned regime \cite{Nattermann, ChauvePRB, Kleemann, IntrinsicTatara, ThiavilleEPL, ExtrinsicTatara, IntrinsicOno, TataraReview, ExtrinsicNatPhys} and is periodically deformed, creating a time-varying effective U(1) gauge field and thus an oscillating EEF. From an energetic perspective, an emergent inductor stores energy in a magnetic texture under a current \cite{Furuta1, Furuta2}, in contrast to classical inductors, which store energy as the magnetic field under a current \cite{Jackson}.

Soon after the theoretical proposal of emergent inductors \cite{NagaosaJJAP}, experimental studies were launched. The AC impedance $Z(\omega)$, also termed complex resistivity $\rho(\omega)$ (normalized with the sample dimensions), where $\omega$ represents the angular frequency, was extensively investigated for materials that contain a noncollinear magnetic texture \cite{YokouchiNature, KitaoriPNAS, YokouchiArxiv, KitaoriPRB}. Thus far, the imaginary part of the complex resistivity of such magnetic materials divided by $\omega$, i.e., $\Im \rho^{1\omega}(\omega, j_0)/\omega$, has been commonly found to be negligibly small for a weak AC current density, $j_0 e^{i \omega t}$, whereas it is significantly enhanced and becomes detectable for a relatively large AC current density, $j_0 \sim 10^{8}$ A~m$^{-2}$. Here, the complex $\rho^{1\omega}(\omega, j_0)$ is not differential resistivity but defined by the 1$\omega$ Fourier component of the time-varying electric field under $j_0 e^{i \omega t}$ divided by $j_0$. The nonlinearly enhanced $\Im \rho^{1\omega}(\omega, j_0)/\omega$ has been interpreted as emergent inductance. The present authors believe, however, that this interpretation needs to be reconsidered; in particular, several observations seem to be not well explained within the EEF-based inductance scenario. Here, we raise fundamental questions  associated with the interpretation.

\textit{Question  I}. The values of $\Im \rho^{1\omega} (j_0, \omega)/\omega$ reported in experiments \cite{YokouchiNature, KitaoriPNAS, YokouchiArxiv, KitaoriPRB} appear too large to be ascribed to the EEF origin. Linear-response EEF theory includes $\hbar$ and predicts only a small value of $\Im \rho^{1\omega}(\omega)/\omega$, on the order of 10$^{-11}$--10$^{-13}$ $\mu \Omega$~cm~s \cite{NagaosaJJAP, Furuta1}. In contrast, for $j_0 \sim 10^8$ A~m$^{-2}$, the magnitudes of experimentally observed $\Im \rho^{1\omega} (\omega, j_0)/\omega$ at low $\omega$ are $\approx$$-9\times$10$^{-7}$ $\mu \Omega$~cm~s for Gd$_3$Ru$_4$Al$_{12}$ \cite{YokouchiNature}, $\approx $$-4\times$10$^{-4}$ $\mu \Omega$~cm~s for YMn$_6$Sn$_{6}$ \cite{KitaoriPNAS}, and $\approx$$-3\times$10$^{-3}$ $\mu \Omega$~cm~s for FeSn$_2$ \cite{YokouchiArxiv}. Although the experimental reports emphasize that these values are observed in the nonlinear regime, understanding such gigantic responses within the EEF framework is nontrivial even if the EEF beyond the linear-response regime is considered. Furthermore, the sign of the reported $\Im \rho^{1\omega} (\omega, j_0)$ is negative in most cases. The negative $\Im \rho^{1\omega} (\omega, j_0)/\omega$ was interpreted as the inductance being negative, but the theory of dynamical systems concludes that unstable behavior occurs when the coefficient of the time derivative of the electric current, ${\rm d} I/{\rm d} t$, is negative, contradicting the experimental observations \cite{Textbook} (for details, see Supplemental Materials \cite{Supp}).

\textit{Question II}. The $\omega$ dependence of the nonlinear $\Im \rho^{1\omega} (\omega, j_0)/\omega$ reported thus far exhibits a cutoff frequency as low as $\sim$1--10 kHz for micrometer-sized fabricated Gd$_3$Ru$_4$Al$_{12}$ and YMn$_6$Sn$_{6}$ \cite{YokouchiNature, KitaoriPNAS} and $\sim$0.1 kHz for needle-like bulk FeSn$_2$ \cite{YokouchiArxiv}. These results were interpreted as indicating that the magnetic texture under consideration has slow dynamics, but the experimentally obtained resonance frequency of a ferromagnetic domain wall of $\sim$10$^7$ Hz should be noted \cite{SaitoNature}. Currently, there is no understanding of why such extremely slow dynamics are ubiquitously observed in the study of nonlinear emergent inductance. 

\textit{Question III}. In Gd$_3$Ru$_4$Al$_{12}$ \cite{YokouchiNature} and YMn$_6$Sn$_{6}$ \cite{KitaoriPNAS}, $\Im \rho^{1\omega} (\omega, j_0)/\omega$ exhibits complicated magnetic-field dependence in response to the successive magnetic phase transitions. Although these observations reveal a considerable correlation between $\Im \rho^{1\omega} (\omega, j_0)/\omega$ and the magnetic phases, the complicated behavior has not yet been well substantiated in terms of EEF.

\begin{figure*}
\includegraphics{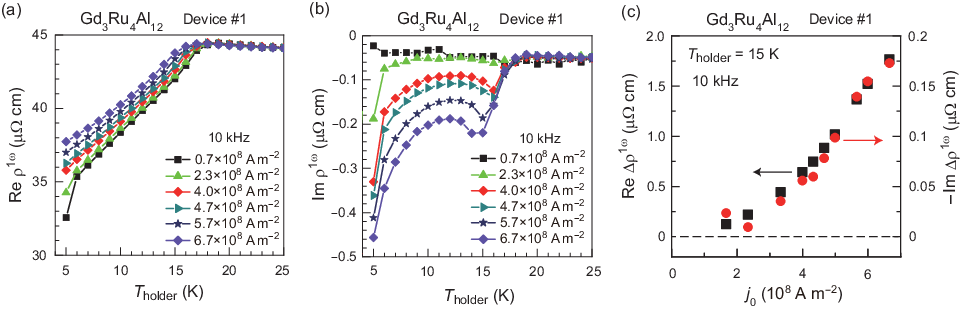}
\caption{\label{Gd3Ru4Al12} Nonlinear $\rho^{1\omega}$ in Gd$_3$Ru$_4$Al$_{12}$. (a, b) $\Re \rho^{1\omega}$--$T$ (a) and $\Im \rho^{1\omega}$--$T$ (b) profiles under various AC current densities, $j_0 e^{i \omega t}$, at $\omega/(2\pi) = 10$ kHz. (c) Current-density dependence of complex $\Delta \rho^{1\omega} (j_0) = \rho^{1\omega} (j_0) - \rho^{1\omega} (0.7\times10^8 \; {\rm A~m^{-2}})$ at 15 K. The panels are constructed from the raw data published in the literature \cite{YokouchiNature}.
}
\end{figure*}

In exploring clues to answer questions I--III, we realize that the magnitude of the nonlinear enhancement of $\Re \rho^{1\omega} (\omega, j_0)$ is larger than that of $\Im \rho^{1\omega} (\omega, j_0)$; i.e., regarding the nonlinear part, $\Re \Delta \rho^{1\omega} > \Im \Delta \rho^{1\omega}$ holds, where $\Delta \rho^{1\omega} (\omega, j_0) \equiv \rho^{1\omega} (\omega, j_0) - \rho_0 (\omega)$, with $\rho_0(\omega)$ representing a linear-response complex value. For Gd$_3$Ru$_4$Al$_{12}$, for instance, the $j_0$-dependent $\Re \rho^{1\omega}(j_0)$--$T$ and $\Im \rho^{1\omega}(j_0)$--$T$ profiles in Figs.~\ref{Gd3Ru4Al12}(a) and (b), respectively, were reported \cite{YokouchiNature}. A static temperature increase cannot explain the pronounced variations in $\Im \rho^{1\omega} (j_0)$ at 10 kHz, and on this basis, the nonlinear $\Im \rho^{1\omega}(j_0)$ was interpreted as originating from the current-induced EEF. However, the nonlinear part, which is here defined as $\Delta \rho^{1\omega} (j_0) \equiv \rho^{1\omega} (j_0) - \rho^{1\omega} (0.7\times 10^8 \; {\rm A~m^{-2}}$), exhibits a $\Re \Delta \rho^{1\omega}$ that is approximately 10 times larger than $\Im \Delta \rho^{1\omega}$ for all $j_0$ values [Fig.~\ref{Gd3Ru4Al12}(c)]. $\Re \rho \gg \Im \rho$ at low $\omega$ is a characteristic of dissipative responses of a resistor, and therefore, the nonlinear impedance $\Delta \rho^{1\omega} (j_0)$ is resistor-like. This perspective is in contrast to the nonlinear inductance mechanism, which assumes nondissipative characteristics, i.e., $\Im \Delta \rho^{1\omega} \gg \Re \Delta \rho^{1\omega}$ at low $\omega$. The observation of $\Re \rho \gg \Im \rho$ at low $\omega$ implies that the nonlinear impedance observed in Gd$_3$Ru$_4$Al$_{12}$ represent not a manifestation of an nonlinear emergent inductor but a manifestation of a nonlinear resistor. Thus, it appears more reasonable to regard the $\Im \Delta \rho^{1\omega}(j_0)$ under an AC current as a delayed response of the much larger $\Re \Delta \rho^{1\omega}(j_0)$, rather than the emergent inductance.

This notice led us to reconsider the unresolved issues regarding nonlinear $\Im \Delta\rho^{1\omega}$ from a perspective of a dissipative mechanism. In general, when nonlinear electrical responses under a large current are examined, the impact of Joule heating must be considered. In AC impedance measurements, the sample temperature is time-varying with a $2\omega$ modulation as a result of the time-varying current. Thus, the situation is more complicated than DC measurements, in which Joule heating only results in a static temperature increase. Such time-varying heating may cause a nonlinear $\Im \Delta\rho^{1\omega}$. While its impact has been discussed in the context of detecting a superconducting transition \cite{HeatingZ}, to the best of the authors' knowledge, the impact of time-varying Joule heating on the nonlinearly enhanced $\Im \rho^{1\omega}$ has not been discussed in the experimental studies on emergent inductance. This finding motivated us to scrutinize whether the nonlinear $\Im \rho^{1\omega}$ caused by AC Joule heating is truly negligible in the reported results \cite{YokouchiNature, KitaoriPNAS, YokouchiArxiv}.

The paper is organized as follows: In Sec.~II, we construct a Joule-heating model up to the power-linear order and derive the expressions of the Joule-heating-induced AC electrical response. In Sec.~III, we show the experimental observations of the Joule-heating-induced AC electrical response in CuIr$_2$S$_4$, which contains no magnetic order, and verify our Joule-heating model. Section IV is devoted to reconsidering the nonlinear emergent inductance previously reported and showing that the Joule-heating-induced AC electrical response approximately reproduces the temperature and magnetic field dependences, cutoff frequency, and magnitude of the $\Im \rho/\omega$. In Sec.~V, we consider the coexistence of dissipative and nondissipative mechanisms in impedance and argue that a nonlinear low-frequency regime should be avoided when exploring nondissipative signals in conducting materials. Section VI summarizes and concludes the paper.

\section{II. Joule heating model for the nonlinear AC electrical response}

\subsection{A. Model construction}
Below, we clarify $\rho(\omega, j_0)$ characteristics when a large AC current is applied to the sample such that the effect of time-varying Joule heating is not negligible; that is, we consider a sample in contact with a heat bath of temperature $T_0$ and derive the voltage responses under the time-varying sample temperature, $T(t)$, due to Joule heating. For simplicity, we disregard the temperature gradient within the sample. To analytically solve this problem, we make the following assumptions that appear to be physically reasonable.

\textit{Assumption $\#$1}. The instantaneous voltage drop in the time-varying self-heated sample is given by:
\begin{equation}
\label{ohm}
V(t) = R_0 \bigl( T(t) \bigr) I(t),
\end{equation}
where $R_0(T)$ denotes the linear-response resistance at $T$. In this model, only this resistive mechanism is considered for the relationship between the voltage and current. In other words, neither inductive nor capacitive mechanisms are considered.

\textit{Assumption $\#$2}. We treat the temperature increase of the sample, $\Delta T(t) = T(t)-T_0$, with respect to the AC power input with angular frequency $\Omega$, $P(t) = \Re ( P_0e^{i\Omega t} )$, as being within the power-in-linear-response regime. Thus, we consider the lowest-order nonlinear response to the AC current input. By introducing a complex response function, $\chi^*(\Omega, T_0) = \chi'(\Omega, T_0) - i\chi''(\Omega, T_0)$, $\Delta T(t)$ is given by:
\begin{align}
\label{DeltaT}
\Delta T(t) &= \Re \bigl[ \chi^*(\Omega, T_0) P_0e^{i\Omega t} \bigr] \nonumber \\
&= P_0 \bigl[ \chi'(\Omega, T_0) \cos \Omega t + \chi''(\Omega, T_0) \sin \Omega t \bigr],
\end{align}
which is the expression for a cosine-wave power input, $P(t) = P_0 \cos \Omega t$. By definition, in the DC limit, $\chi'(\Omega, T_0)$ approaches a finite value, $\chi_0(T_0)$, and $\chi''(\Omega, T_0)$ approaches zero; i.e., we impose $\lim_{\Omega \to 0} \chi'(\Omega, T_0) = \chi_0(T_0)$ and $\lim_{\Omega \to 0} \chi''(\Omega, T_0) = 0$. Note that $\chi^*$ is determined by the heat capacitance of the system and the heat conduction to the heat bath; thus, $\chi^*$ depends on the volume and geometry of the sample, the details of the thermal contacts, etc.

\begin{figure}
\includegraphics{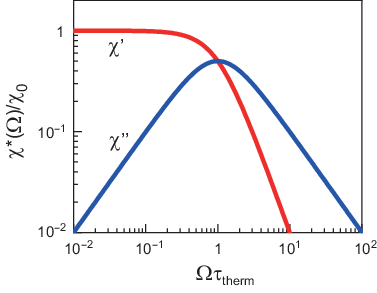}
\caption{\label{Chi} Example of the response function, $\chi^*(\Omega) = \chi' - i\chi''$, which describes the temperature change in response to the power input. The functional form of Eq.~(\ref{Debye}) with $\alpha = 0$ is displayed on a double logarithmic scale.}
\end{figure}

\textit{Assumption $\#$3}. A system has a nonzero thermal-response time, $\tau_{\rm therm}$ $(>0)$, which is a phenomenon known as thermal relaxation. Thus, under an AC power input, the thermal response of the sample is more or less delayed (i.e., $\chi'' > 0$), and correspondingly, $\chi^*(\Omega, T_0)$ has a $\Omega$ dependence with a cutoff frequency of $\approx 1/(2\pi \tau_{\rm therm})$. The form of $\chi^*(\Omega, T_0)$ can be approximately captured by a polydispersive Cole-Cole-type response:
\begin{equation}
\label{Debye}
\chi^*(\Omega, T_0)=\frac{\chi_0(T_0)}{1+ \bigl( i\Omega\tau_{\rm therm}(T_0) \bigr)^{1-\alpha}},
\end{equation}
where $\alpha$ represents the polydispersivity. For the readers' reference, we display the functional form of Eq.~(\ref{Debye}) for the case of a monodispersive relaxation, $\alpha =0$, in Fig.~\ref{Chi}. The details of $\chi^*$ depend on the system, as mentioned above. Nevertheless, the only important feature in the following discussion is that $\chi^*(\Omega, T_0)$ has a cutoff frequency determined by the thermal-response dynamics and satisfies $\lim_{\Omega \to 0} \chi'(\Omega, T_0) = \chi_0(T_0)$ and $\lim_{\Omega \to 0} \chi''(\Omega, T_0) = 0$. Equation (\ref{Debye}) is an example of a function that satisfies these characteristics.

\textit{Assumption $\#$4}. $\Delta T(t)$ is so small that the time-dependent resistance, $R\bigl( T(t) \bigr)$, is well approximated by:
\begin{align}
\label{Rchange}
R\bigl( T(t) \bigr) &= R_0(T_0) + \Delta R_0 \bigl( \Delta T(t) \bigr) \nonumber \\
&\approx R_0(T_0) + \frac{{\rm d} R_0(T_0)}{{\rm d} T}\Delta T(t),
\end{align}
where ${\rm d} R_0(T_0)/{\rm d} T \equiv {\rm d}R_0/{\rm d} T\rvert_{T=T_0}$.

\subsection{B. Derivation of the nonlinear AC electrical response}

With the above assumptions, one can analytically derive $P(t)$, $\Delta T(t)$, $R(t)$, and $V(t)$ in sequence. We consider the situation in which a sine-wave AC current with angular frequency $\omega$, $I(t) = I_0\sin \omega t$, is applied to the sample. Hence, the time-varying power input, $P(t)$, is given by:
\begin{align}
\label{Power}
P(t) &\approx R_{\rm 2probe}(T_0)(I_0 \sin \omega t)^2 \nonumber \\ 
&= P_0 \frac{1-\cos 2\omega t}{2 }\nonumber \\
&= \frac{P_0}{2} - \frac{P_0}{2}\Re (e^{i2\omega t} ),
\end{align}
where $R_{\rm 2probe}$ denotes the two-probe resistance including the contact resistance $R_{\rm contact}$ and $P_0(I_0, T_0) \equiv R_{\rm 2probe}(T_0)I_0^2 = \bigl( R_0(T_0) + R_{\rm contact}(T_0) \bigr) I_0^2$. Here, we disregard the deviation from Eq.~(\ref{Power}) due to the time-varying, two-probe resistance.
By combining this result with Eq.~(\ref{DeltaT}) and further considering Eqs.~(\ref{Rchange}) and (\ref{ohm}) in sequence, one obtains:
\begin{align}
\Delta T(t) &\approx \frac{\chi_0(T_0) P_0}{2} - \frac{P_0}{2} \Re \Bigl[ \chi^*(2\omega, T_0) e^{i2\omega t} \Bigr] \nonumber \\
&= \frac{\chi_0(T_0) P_0}{2} - \frac{P_0}{2} \Bigl[ \chi'(2\omega, T_0) \cos 2\omega t + \chi''(2\omega, T_0) \sin 2\omega t \Bigr]. \nonumber \\
\therefore R(t) &\approx R_0(T_0) + \frac{{\rm d} R_0(T_0)}{{\rm d} T} \left\{ \frac{\chi_0(T_0) P_0}{2} - \frac{P_0}{2} \Bigl[ \chi'(2\omega, T_0) \cos 2\omega t + \chi''(2\omega, T_0) \sin 2\omega t \Bigr] \right\}. \nonumber \\
\therefore V(t) &\approx R_0(T_0) I_0 \sin \omega t \nonumber \\
&+ \frac{{\rm d} R_0(T_0)}{{\rm d} T} \left \{ \frac{\chi_0(T_0) P_0}{2} - \frac{P_0}{2} \Bigl[ \chi'(2\omega, T_0) \cos 2\omega t + \chi''(2\omega, T_0) \sin 2\omega t \Bigr] \right \} I_0 \sin \omega t. \nonumber \\
\label{Fourier}
\therefore \frac{V(t)}{I_0} &\approx R_0(T_0)\sin \omega t + \frac{{\rm d} R_0(T_0)}{{\rm d} T} \frac{\chi_0(T_0) P_0}{2}\sin \omega t \nonumber \\
&+ \frac{{\rm d} R_0(T_0)}{{\rm d} T} \frac{P_0}{4} \Bigl[\chi'(2\omega, T_0) \sin \omega t - \chi''(2\omega, T_0) \cos \omega t - \chi'(2\omega, T_0) \sin 3\omega t + \chi''(2\omega, T_0) \cos 3 \omega t \Bigr]. 
\end{align}
Thus, for an input of $I(t) = I_0 \sin \omega t$, the Fourier series of $V(t)/I_0$ has been analytically derived up to the 3$\omega$ components. Following previous studies \cite{YokouchiNature, KitaoriPNAS, YokouchiArxiv}, we introduce $Z^{n\omega}$ to describe the resulting in-phase and out-of-phase electrical responses of the $n \omega$ components ($n = 1, 3$); i.e., $\Re Z^{n\omega} \equiv \Re \mathcal{V}^{n\omega}(\omega, I_0)/I_0$ and $\Im Z^{n\omega} \equiv \Im \mathcal{V}^{n\omega}(\omega, I_0)/I_0$, where $\mathcal{V}^{n\omega}(\omega, I_0)$ is the $n \omega$ Fourier component of $V(t)$ under an AC current with $\omega$. The real and imaginary parts of $Z^{n\omega}$ are given by: 
\begin{align}
\label{Re1w}
\Re Z^{1\omega}(\omega, I_0, T_0) - R_0(T_0) &= \frac{{\rm d} R_0(T_0)}{{\rm d} T} \frac{P_0(I_0, T_0)}{4} \bigl[ 2\chi_0(T_0) + \chi'(2\omega, T_0) \bigr], \\
\label{Im1w}
\Im Z^{1\omega}(\omega, I_0, T_0) &= - \frac{{\rm d} R_0(T_0)}{{\rm d} T} \frac{P_0(I_0, T_0)}{4} \chi''(2\omega, T_0), \\
\label{Re3w}
\Re Z^{3\omega}(\omega, I_0, T_0) &= - \frac{{\rm d} R_0(T_0)}{{\rm d} T} \frac{P_0(I_0, T_0)}{4}\chi'(2\omega, T_0), \\
\label{Im3w}
\Im Z^{3\omega}(\omega, I_0, T_0) &= \frac{{\rm d} R_0(T_0)}{{\rm d} T} \frac{P_0(I_0, T_0)}{4} \chi''(2\omega, T_0).
\end{align}
Note that the right-hand sides of Eqs.~(\ref{Re1w})--(\ref{Im3w}) are $I_0$ dependent via $P_0(I_0, T_0) = R_{\rm 2probe}(T_0)I_0^2$, and thus, they represent nonlinear responses. The emergence of these nonlinear terms can be qualitatively understood as follows: Under an AC current with angular frequency $\omega$, the sample temperature and resistance are time-varying, with a $2\omega$ modulation; the $2\omega$ resistance modulation couples with the AC current with $\omega$, generating additional output voltage modulations of both $\omega$ and $3\omega$. Thus, $\Im Z^{1\omega}$ and $\Im Z^{3\omega}$ appear due to a delay of the thermal response (i.e., $\chi''$). Note again that Eqs.~(\ref{Re1w})--(\ref{Im3w}) are valid only when variations of ${\rm d}R_0/{\rm d} T$ and $\chi^*$ under a Joule-heating-induced temperature oscillation are negligible. This condition becomes less likely to be satisfied at and near a phase transition: in such a critical region, ${\rm d}R_0/{\rm d} T$ and $\chi^*$ may pronouncedly change as a function of temperature.

\subsection{C. Characteristics of the Joule-heating-induced AC electrical response}

Having derived the expressions of the Joule-heating-induced AC electrical response up to the power-linear order, Eqs.~(\ref{Re1w})--(\ref{Im3w}), we now discuss their characteristics. Below we use $T$ to denote the sample-holder temperature, which corresponds to $T_0$ of the previous sections. Because the physical quantities also depend on the magnetic field, $H$, the function arguments shall include $H$ in addition to $T$. In that case, ${\rm d}R_0/{\rm d}T$ should be read as ${\partial}R_0/{\partial}T$.

First, $\Im Z^{1\omega} = -\Im Z^{3\omega}$ holds.

Second, the expressions of $\Im Z^{1\omega}$ $(= -\Im Z^{3\omega})$ and $\Re Z^{3\omega}$ involve $P_0$$\times$${\rm d}R_0/{\rm d}T$ [Eqs.~(\ref{Im1w})--(\ref{Im3w})]. Thus, the variations in $\Im Z^{1\omega/3\omega}$ when changing temperature or magnetic field are correlated with those in ${\rm d}R_0/{\rm d}T$. Since $P_0 = \bigl( R_0(T) + R_{\rm contact}(T) \bigr) I_0^2$, whether ${\rm d}R_0/{\rm d}T$ or $R_0$$\times$${\rm d}R_0/{\rm d}T$ better describes the $\Im Z^{1\omega/3\omega}$ variations depends on whether the Joule heating is dominant in the bulk or at the contacts of the current electrodes. If the Joule heat is produced exclusively in the bulk [i.e., $P_0 \approx R_0(T)I_0^2$], then $\Im Z^{1\omega/3\omega}$ variations would scale with $R_0$$\times$${\rm d} R_0/{\rm d} T$. If the Joule heat is generated mainly at the contacts of the current electrodes [i.e., $ P_0 \approx R_{\rm contact}(T)I_0^2$] and if $R_{\rm contact}(T)$ only weakly depends on $T$, then $\Im Z^{1\omega/3\omega}$ variations would scale with ${\rm d} R_0/{\rm d} T$. However, the scaling should only be qualitative because the $\chi^*(\omega, T)$, which scales with the heat conductance and inversely scales with the heat capacitance, is also temperature and magnetic field dependent.

Third, the Joule-heating-induced AC electrical response has a cutoff frequency $\omega_c$ that is determined by the thermal relaxation time of the sample: $\omega_c \approx 1/(2\tau_{\rm therm})$. The thermal relaxation time depends on the system details, such as the sample dimensions and thermal contacts. For a millimeter-sized bulk sample, the typical value of $\tau_{\rm therm}$ can be as long as $\sim$10$^{-1}$--10$^{-2}$ s \cite{OikeNatPhys}, which leads to $\sim$10$^0$--10$^1$ Hz for the cutoff frequency of $\chi^*(\omega)$. For an exfoliated thin plate with a submicrometer thickness, the typical thermal relaxation time is $\sim$10$^{-6}$--10$^{-3}$ s (depending on the thermal conductivity of the substrate) \cite{OikeSciAdv}, which leads to $\sim$10$^2$--10$^5$ Hz for the cutoff frequency. As shown in Sec.~III, our microfabricated CuIr$_2$S$_4$ and bulk MoTe$_2$ exhibit cutoff frequencies of 20 kHz and 1 Hz, respectively. Thus, in general, the cutoff frequency of the Joule-heating-induced AC electrical response can be much lower than the resonance frequency of the magnetic texture under consideration.

\begin{figure*}
\includegraphics{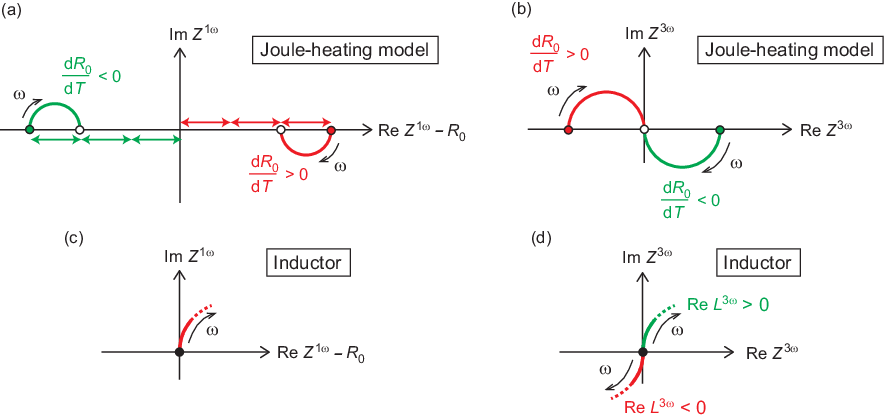}
\caption{\label{ColeCole} Schematic Cole-Cole representation of the AC electrical response. (a, b) Cole-Cole representations of $Z^{1\omega}$ (a) and $Z^{3\omega}$ (b) under Joule heating. $R_0$ represents the DC linear-response impedance. Schematics are drawn for both cases of ${\rm d}R_0/{\rm d}T > 0$ and ${\rm d}R_0/{\rm d}T < 0$. (c, d) Cole-Cole representations of $Z^{1\omega}$ (c) and $Z^{3\omega}$ (d) of a nonlinear inductor element.}
\end{figure*}

Fourth, the following relations hold at low frequencies, $\omega \ll \omega_c = 1/(2\tau_{\rm therm}$):
\begin{align}
\label{1w-limit}
\left| \frac{\Im Z^{1\omega}}{\Re Z^{1\omega}-R_0} \right| = \frac{\chi''(2\omega, T_0)}{ 2\chi_0(T_0) +\chi'(2\omega, T_0) } \ll 1 \hspace{0.5cm} \rm for \hspace{0.2cm} \omega \ll \omega_c.\\
\label{3w-limit}
\left| \frac{\Im Z^{3\omega}}{\Re Z^{3\omega}} \right| = \frac{\chi''(2\omega, T_0)}{\chi'(2\omega, T_0)} \ll 1 \hspace{0.5cm} \rm for \hspace{0.2cm} \omega \ll \omega_c.
\end{align}
These equations indicate that at low $\omega$, the nonlinearly induced change from the linear-response impedance occurs mainly in the real part, rather than in the imaginary part. These characteristics are also evident in the Cole-Cole representation [Figs.~\ref{ColeCole}(a) and (b)], in which $\Re Z^{1\omega}-R_0$ is adopted for the real axis for clarity. In the Cole-Cole representation of $Z^{3\omega}$, the $\omega$-evolving trajectory starts somewhere on the real axis at DC and converges to the origin at high $\omega$ [Fig.~\ref{ColeCole}(b)]. The behavior shown in Figs.~\ref{ColeCole}(a) and (b) indicates that the Joule-heating-induced AC electrical response has dissipative characteristics. In contrast, when nonlinear $Z^{1\omega}$ and $Z^{3\omega}$ are caused by the nonlinear inductance, the nonlinear change should exclusively appear in the imaginary part; that is, at low $\omega$, $\left| \Im Z^{1\omega} \right| \gg \left| \Re Z^{1\omega}-R_0 \right| $ and $\left| \Im Z^{3\omega} \right| \gg \left| \Re Z^{3\omega} \right| $ should hold. In the Cole-Cole representation, these nondissipative characteristics are observed as shown in Figs.~\ref{ColeCole}(c) and (d). In particular, the $\omega$-evolving trajectory of $Z^{3\omega}$ should start from the origin at DC [Fig.~\ref{ColeCole}(d)], which is distinctly different from the $Z^{3\omega}$ caused by Joule heating [Fig.~\ref{ColeCole}(b)]. Thus, the Cole-Cole representation provides key insight into whether the observed nonlinear AC electrical response has dissipative or nondissipative characteristics.

\section{III. Experiments}

\begin{figure*}
\includegraphics{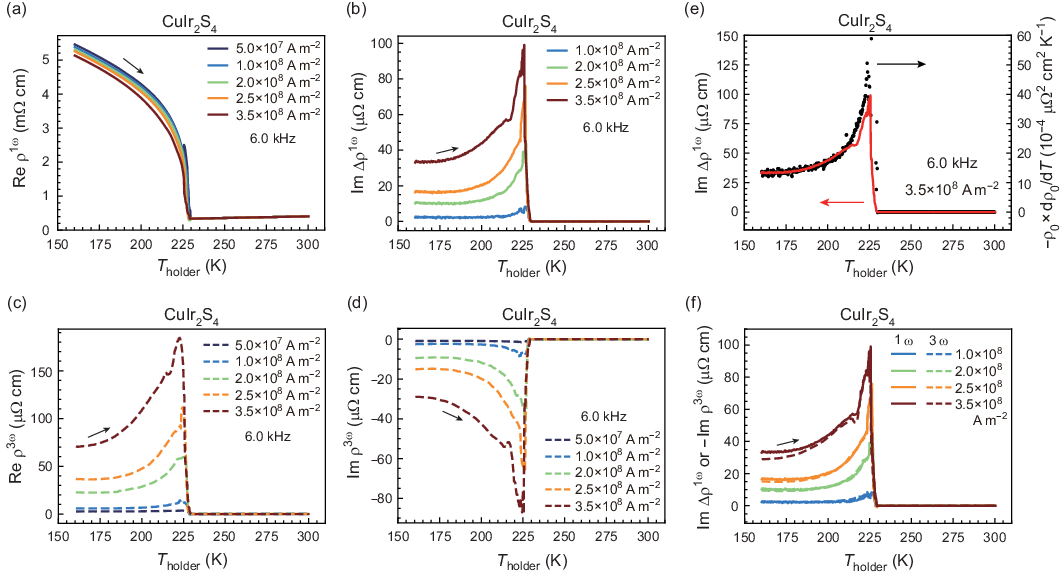}
\caption{\label{CuIr2S4} Temperature dependence of the AC electrical response of CuIr$_2$S$_4$ measured at various AC current densities. (a) $\Re \rho^{1\omega}$, (b) $\Im \Delta \rho^{1\omega}$, (c) $\Re \rho^{3\omega}$, and (d) $\Im \rho^{3\omega}$. The $\Im \Delta \rho^{1\omega}$ shown in (b) is defined as $\Im \Delta \rho^{1\omega} \equiv \Im \rho^{1\omega}(j_0) - \Im \rho^{1\omega}(5.0\times10^{7}$ A~m$^{-2}$). (e) Comparison between $\Im \Delta \rho^{1\omega}$ and $-\rho_0$$\times$${\rm d}\rho_0/{\rm d}T$. The data were recorded in the heating process. (f) Comparison between $\Im \Delta \rho^{1\omega}$ and $-\Im \rho^{3\omega}$. }
\end{figure*}

To experimentally test the Joule heating model, we measured the nonlinear AC electrical response for two conducting systems: a microfabricated CuIr$_2$S$_4$ crystal (sample dimensions of 20$\times$4$\times$1 $\mu$m$^3$) and a bulk 1$T$'-MoTe$_2$ crystal (approximately, 1.3$\times$0.7$\times$0.14 mm$^3$); images of the two samples are shown in the APPENDIX [Figs.~\ref{Photo}(a) and (b)]. These materials exhibit no magnetic order, and a current-induced spin dynamics contribution to the AC electrical response can therefore be ruled out. In both materials, we find a good agreement between the experimental results and expected results for the Joule-heating-induced AC electrical response. For readability, only results for the microfabricated CuIr$_2$S$_4$ are shown in the main text. The results for the bulk MoTe$_2$ are shown in the APPENDIX (Figs.~\ref{MoTe2} and \ref{MoTe2_freq}).

CuIr$_2$S$_4$ shows a first-order metal-insulator transition at $T_{\rm c} \approx$ 230 K \cite{CuIr2S4_Furubayashi, CuIr2S4_Cheong}. This material is paramagnetic and metallic (i.e., ${\rm d}\rho_0/{\rm d}T > 0$ with $\rho_0$ representing a linear-response DC value) above $T_{\rm c}$, whereas it is nonmagnetic and semiconducting (i.e., ${\rm d}\rho_0/{\rm d}T < 0$) below $T_{\rm c}$. Figures \ref{CuIr2S4}(a), (b), (c), and (d) display the temperature dependences of $\Re \rho^{1\omega}$, $\Im \Delta \rho^{1\omega}$, $\Re \rho ^{3\omega}$, and $\Im \rho^{3\omega}$ at $\omega$/2$\pi = 6$ kHz, respectively, measured at various AC current densities. Here, we display $\Im \Delta \rho^{1\omega} \equiv \Im \rho^{1\omega}(j_0) - \Im \rho^{1\omega}(5.0\times10^{7}$ A~m$^{-2}$) only for $\Im \rho^{1\omega}$ to subtract the contribution of the nonnegligible linear-response background; for the raw data of $\Im \rho^{1\omega}$, see the APPENDIX [Fig.~\ref{Photo}(c)]. In contrast, for $\rho^{3\omega}$, we discover that the background signal is not significant, so we display the raw data.

In the $\Re \rho ^{1\omega}$--$T$ profile [Fig.~\ref{CuIr2S4}(a)], the apparent transition temperature clearly decreases with increasing current, indicating that the sample temperature is elevated from the sample-holder temperature, $T_{\rm holder}$, by Joule heating. In the insulating phase, the four-probe resistance is 30--80 $\Omega$ within 225--160 K, whereas the contact resistance is $\approx$20 $\Omega$. Thus, the Joule heat is assumed to be generated mainly in the bulk, rather than at the contacts of the current electrodes.

Figures~\ref{CuIr2S4}(b)--(f) show characteristic features consistent with the Joule heating model. First, in the insulating phase, $\Im \Delta \rho^{1\omega}$, $\Re \rho^{3\omega}$ and $\Im \rho^{3\omega}$ nonlinearly emerge as the AC current density increases [Figs.~\ref{CuIr2S4}(b)--(d)]. The signs of these quantities are consistent with Eqs.~(\ref{Im1w})--(\ref{Im3w}) for the case of ${\rm d}\rho_0/{\rm d}T<0$. Second, the $\Im \Delta\rho^{1\omega}$--$T$ profile agrees well with the $(-\rho_0$$\times$${\rm d}\rho_0/{\rm d}T)$--$T$ profile [Fig.~\ref{CuIr2S4}(e)], which is consistent with the results expected when Joule heating occurs mainly in the bulk. Third, $\Im \Delta \rho^{1\omega} = -\Im \rho^{3\omega}$ is well satisfied [Fig.~\ref{CuIr2S4}(f)].

\begin{figure*}
\includegraphics{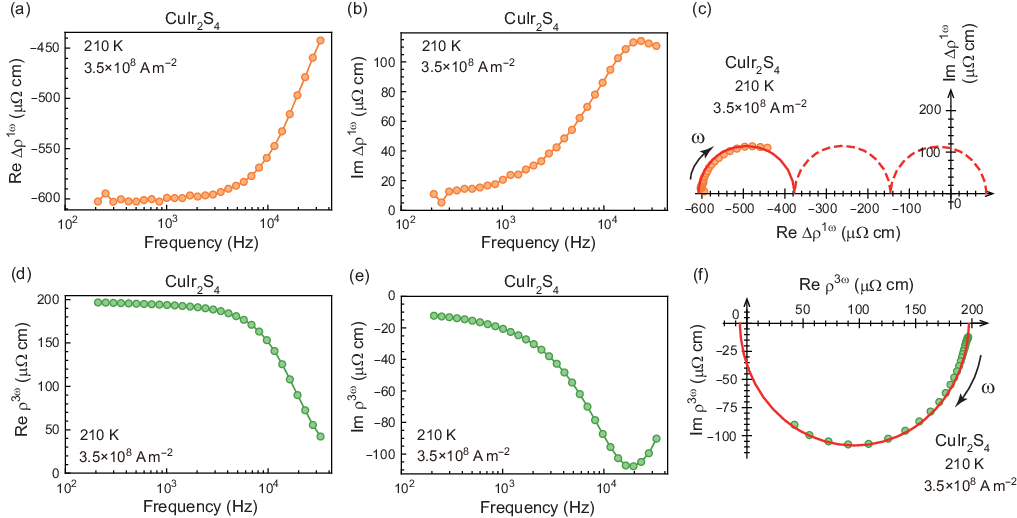}
\caption{\label{CuIr2S4_freq} Frequency dependence of the nonlinear AC electrical response of CuIr$_2$S$_4$ at 210 K. (a) $\Re \Delta \rho^{1\omega}$, (b) $\Im \Delta\rho^{1\omega}$, and (c) Cole-Cole representation of $\Delta \rho^{1\omega}$. $\Delta \rho^{1\omega}$ is defined as $\Delta \rho^{1\omega} \equiv \rho^{1\omega}(j_0) - \rho^{1\omega}(5.0\times10^{7}$ A~m$^{-2}$). (d) $\Re \rho^{3\omega}$, (e) $\Im \rho^{3\omega}$, and (f) Cole-Cole representation of $\rho^{3\omega}$. The data were recorded at $j_0 = 3.5\times10^8$ A~m$^{-2}$.
}
\end{figure*}

The frequency dependences and Cole-Cole representations of $\Delta \rho^{1\omega}$ and $\rho^{3\omega}$ at 210 K are shown in Figs.~\ref{CuIr2S4_freq}(a)--(f). The Cole-Cole representations [Figs.~\ref{CuIr2S4_freq}(c) and (f)] are consistent with the predictions of the Joule heating model for the case of ${\rm d}\rho_0/{\rm d}T < 0$ [Figs.~\ref{ColeCole}(a) and (c)]. The lengths of the arc strings are approximately the same for both cases (200 $\mu \Omega$~cm). These observations confirm that the observed $\Delta \rho^{1\omega}$ and $\rho^{3\omega}$ have dissipative characteristics and that the origin of their imaginary parts lies in the delay of the nonlinear real parts. From the frequency dependence, the cutoff frequency in the present device is found to be $\approx$20 kHz. Note that this value is not an intrinsic quantity of the material but should depend on the sample volume, details of the thermal contacts, etc. In the bulk MoTe$_2$, for instance, the cutoff frequency is as low as 1 Hz (see the APPENDIX, Fig.~\ref{MoTe2_freq}), indicating that the sample dimensions are a crucial factor determining the cutoff frequency of the Joule-heating-induced AC electrical response. The cutoff frequency in the microfabricated CuIr$_2$S$_4$ depends on temperature only weakly, except for at the transition point, at which it decreases to 7 kHz. This decrease in the cutoff frequency is ascribed to an apparent increase in the heat capacity at a first-order phase transition.

\begin{figure}
\includegraphics{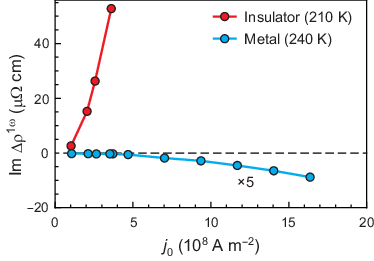}
\caption{\label{IV} Current-density dependence of $\Im \Delta\rho^{1\omega}$ in the insulating and metallic phases. $\Delta \rho^{1\omega}$ is defined as $\Delta \rho^{1\omega} \equiv \Im \rho^{1\omega}(j_0) - \rho^{1\omega}(5.0\times10^{7}$ A~m$^{-2}$).}
\end{figure}

For an applied current density below 5$\times$10$^8$ A~m$^{-2}$, the nonlinear AC electrical response is difficult to detect in the metallic phase above 230 K. To observe the sign of the nonlinear behavior in the metallic phase, we measured the $\Im \Delta \rho^{1\omega}$--$j_0$ profile up to a higher current density, and the results are shown in Fig.~\ref{IV}. The finite nonlinear $\Im \Delta \rho^{1\omega}$ exhibits a detectable magnitude in the metallic phase when $j_0$ exceeds 7$\times10^8$ A~m$^{-2}$, and its sign is negative. Thus, we confirm that the sign of the nonlinear $\Im \Delta\rho^{1\omega}$ is negative in the metallic phase (${\rm d}\rho_0/{\rm d}T > 0$), whereas it is positive in the insulating phase (${\rm d}\rho_0/{\rm d}T < 0$). The relationship between the signs of $\Im \Delta\rho^{1\omega}$ and ${\rm d}\rho_0/{\rm d}T$ is consistent with the Joule heating model.

\section{IV. Reconsidering the nonlinear emergent inductance: answers to questions I--III }

Having experimentally verified the Joule-heating-induced AC electrical response, we examine whether the nonlinear $\Im \rho^{1\omega}(\omega, j_0)$ previously reported \cite{YokouchiNature, KitaoriPNAS, YokouchiArxiv} has the characteristics of the Joule-heating-induced AC electrical response. In particular, we discuss the temperature and magnetic field dependences, cutoff frequency, and magnitude of $\Im \rho^{1\omega}$, and show that the Joule-heating-induced AC electrical response gives answers to questions I--III. We begin by discussing the temperature and magnetic field dependences and answering question III.

\subsection{A. Temperature and magnetic field dependences: answer to question III}

\begin{figure*}
\includegraphics{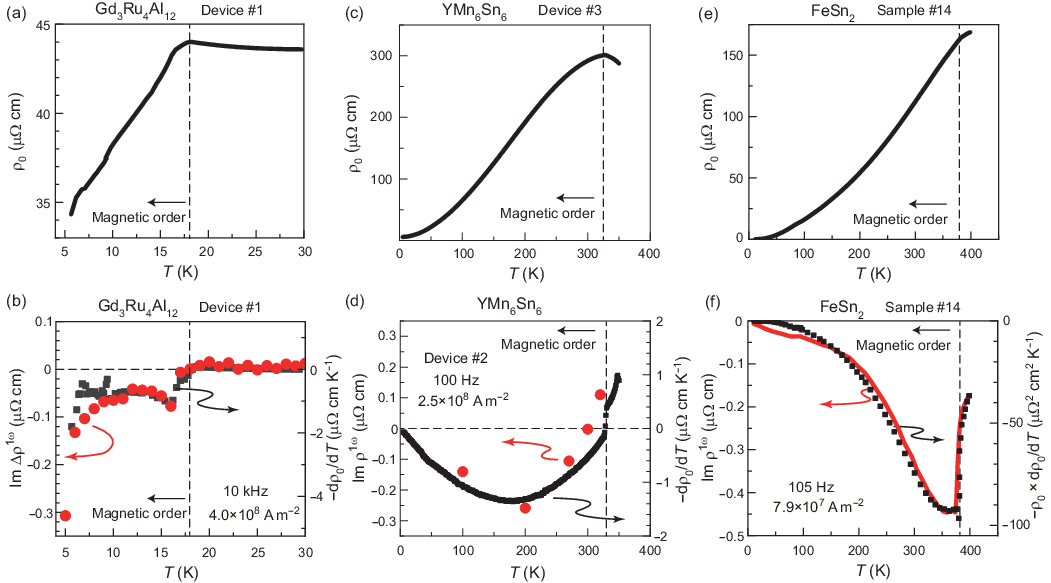}
\caption{\label{Scale} Correlation between nonlinear $\Im \rho(j_0)$ and $-{\rm d}\rho_0/{\rm d} T$ in the reported data. (a, b) $\rho_0$--$T$ profile (a) and $\Im \Delta \rho^{1\omega}(j_0)$--$T$ and $\left( -{\rm d}\rho_0/{\rm d} T \right)$--$T$ profiles (b) of Gd$_3$Ru$_4$Al$_{12}$. (c, d) $\rho_0$--$T$ profile (c) and $\Im \rho^{1\omega} (j_0)$--$T$ and $\left( -{\rm d}\rho_0/{\rm d} T \right)$--$T$ profiles (d) of YMn$_6$Sn$_6$. (e, f) $\rho_0$--$T$ profile (e) and $\Im \rho^{1\omega} (j_0)$--$T$ and $( -\rho_0$$\times$${\rm d}\rho_0/{\rm d} T )$--$T$ profiles (f) of FeSn$_2$. The $\rho_0$--$T$ and $\Im \rho^{1\omega} (j_0)$--$T$ profiles were obtained from the literature \cite{YokouchiNature, KitaoriPNAS, YokouchiArxiv}, and we constructed the ($-{\rm d}\rho_0/{\rm d} T$)--$T$ profiles from the reported data. Note that for YMn$_6$Sn$_6$, the $\Im \rho^{1\omega} (j_0)$--$T$ and $(-{\rm d}\rho_0/{\rm d} T)$--$T$ profiles were collected from different devices.}
\end{figure*}

\begin{figure*}
\includegraphics{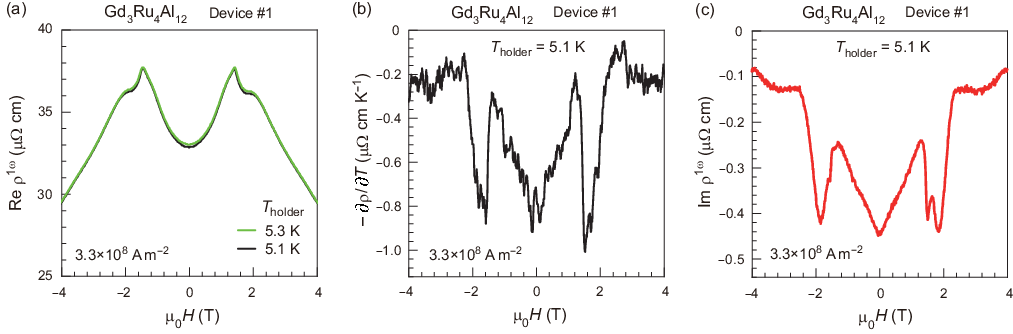}
\caption{\label{Hdep} (a) Magnetic-field dependence of $\Re \rho^{1\omega}(j_0)$ in Gd$_3$Ru$_4$Al$_{12}$ at $T_{\rm holder} = 5.1$ and 5.3 K, measured at a relatively high current density, $j_0 = 3.3\times10^8$ A~m$^{-2}$. (b) Estimated $\big($$-{\partial}\rho(j_0)/{\partial} T$$\big)$--$H$ profile at $j_0 = 3.3\times10^8$ A~m$^{-2}$.
(c) $\Im \rho^{1\omega}(j_0)$--$H$ profile at $j_0 = 3.3\times10^8$ A~m$^{-2}$. The data were collected from the raw data published in the literature \cite{YokouchiNature}.
}
\end{figure*}

As discussed in Sec.~II~C, a main characteristic of the Joule-heating-induced AC electrical response is a correlation between the nonlinearly enhanced $\Im \rho^{1\omega} (\omega, j_0)$ and $-{\rm d}\rho_0/{\rm d}T$. Thus, it is of interest whether the temperature and magnetic field dependences of $\Im \rho^{1\omega} (\omega, j_0)$ are similar to those of $-{\rm d}\rho_0/{\rm d} T$ (or $-\rho_0$$\times$${\rm d}\rho_0/{\rm d} T$).

The comparative results regarding the temperature dependence for Gd$_3$Ru$_4$Al$_{12}$, YMn$_6$Sn$_6$, and FeSn$_2$ are summarized in Fig.~\ref{Scale}. We determine that the $\Im \Delta \rho^{1\omega} (\omega, j_0)$--$T$ profile is similar to the ($-{\rm d}\rho_0/{\rm d} T$)--$T$ profile for Gd$_3$Ru$_4$Al$_{12}$ [Figs.~\ref{Scale}(a) and (b)] \cite{YokouchiNature} and YMn$_6$Sn$_6$ [Figs.~\ref{Scale}(c) and (d)] \cite{KitaoriPNAS} or to the ($-\rho_0$$\times$${\rm d}\rho_0/{\rm d} T$)--$T$ profile for FeSn$_2$ [Figs.~\ref{Scale}(e) and (f)] \cite{YokouchiArxiv}. In YMn$_6$Sn$_6$, notably, the $\Im \Delta \rho^{1\omega} (\omega, j_0)$--$T$ and ($-{\rm d}\rho_0/{\rm d} T$)--$T$ profiles commonly exhibit a sign change.

Figure \ref{Hdep} shows the comparative results regarding the magnetic field dependence for Gd$_3$Ru$_4$Al$_{12}$. The $\Re \rho^{1\omega}(j_0)$--$H$ profiles at $T_{\rm holder} = 5.1$ and 5.3 K were measured at a current density of $j_0 = 3.3\times10^8$ A~m$^{-2}$ [Fig.~\ref{Hdep}(a)]. We estimate the ${\partial}\rho(j_0)/{\partial} T$--$H$ profile at this current density by simply taking the difference between the two sets of data divided by the small $T$ increment, 0.2 K, as shown in Fig.~\ref{Hdep}(b). The $\Im \rho^{1\omega}(j_0)$--$H$ profile at $T_{\rm holder} = 5.1$ K measured at the same current density is shown in Fig.~\ref{Hdep}(c). We determine that the magnetic field dependences shown in Figs.~\ref{Hdep}(b) and (c) are similar. Thus, we conclude that the complicated behavior of $\Im \rho^{1\omega}(j_0)$--$H$ profile originates from that of the $\big($$-{\partial}\rho(j_0)/{\partial} T\big)$--$H$ profile \cite{Note}. This is the answer to question III. 

Because $\Im \rho^{1\omega}(j_0)$ and $-{\partial}\rho_0/{\partial} T$ $\big($or $-{\partial}\rho(j_0)/{\partial} T$$\big)$ show similar temperature and magnetic field dependences, there is a considerable correlation between the two quantities. This correlation appears difficult to understand in terms of the EEF, which is determined by the spin dynamics and does not involve the $T$ derivative of the scattering rate of the charge carrier.

\subsection{B. Cutoff frequency: answer to question II}

As mentioned in Sec.~I, the cutoff frequency of the nonlinear $\Im \rho^{1\omega}$ is $\approx$20 kHz for the microfabricated Gd$_3$Ru$_4$Al$_{12}$ \cite{YokouchiNature}, $\approx$1 kHz for the microfabricated YMn$_6$Sn$_{6}$ \cite{KitaoriPNAS}, and $\approx$0.1 kHz for the needle-like bulk FeSn$_2$ \cite{YokouchiArxiv}. These values and the corresponding sample volumes are shown in Table I, and they are within the range of those observed for the present microfabricated CuIr$_2$S$_4$ device (the cutoff frequency is $\approx$20 kHz and the sample volume is $\approx$80 $\mu$m$^3$) and the bulk MoTe$_2$ crystal (the cutoff frequency is $\approx$1 Hz and the sample volume is $\approx$0.13 mm$^3$). Given that thermal-response dynamics are also affected by details of the thermal contact, the low cutoff frequencies reported previously \cite{YokouchiNature, KitaoriPNAS, YokouchiArxiv} thus appear to be reasonably explained by the time scale of the thermal response. This is the answer to question II.

\subsection{C. Order of magnitude estimate: answer to question I}

\begin{table*}[t]
\caption{Order of magnitude estimate of $\Im \Delta\rho^{1\omega}(\omega, j_0, T)/\omega$ for Gd$_3$Ru$_4$Al$_{12}$ \cite{YokouchiNature}, YMn$_6$Sn$_{6}$ \cite{KitaoriPNAS}, and FeSn$_2$ \cite{YokouchiArxiv}. The parameters used for the calculations are obtained from the literature \cite{YokouchiNature, KitaoriPNAS, YokouchiArxiv} or Fig.~\ref{Gd3Ru4Al12}.}
\centering
\begin{tabular}{ccccc}
\hline
Material & & Gd$_3$Ru$_4$Al$_{12}$ & YMn$_6$Sn$_{6}$ & FeSn$_2$ \\
\hline \hline
Specimen & & Microfabricated & Microfabricated & Bulk \\
Sample-holder temperature, $T_0$ [K] & &15 & 270 & 350 \\
Applied current density, $j_0$ [10$^8$ A~m$^{-2}$] & & 4.0 & 2.5 & 0.8 \\
Sample volume [$\mu$m$^3$] & & 13 & 290 & 8.9$\times10^4$ \\
Observed cutoff frequency, $\omega_c$/2$\pi$ [kHz] & & 20 & 1 & 0.1 \\

$\Re \Delta \rho^{1\omega}(j_0, T)$ at low $\omega$ ($< \omega_c$) [$\mu \Omega$~cm] & & 0.7 & 1.5 & 2.9 \\
\multirow{2}{*}{$\Im \Delta \rho^{1\omega}(\omega, j_0, T)/\omega$ at low $\omega$ ($< \omega_c$) [$\mu \Omega$~cm~s]} & (\textit{Cal.}) & $-2\times$10$^{-6}$ & $-0.8\times$10$^{-4}$ & $-1\times$10$^{-3}$ \\
& (\textit{Exp.}) & $-1.1\times$10$^{-6}$ & $-2.6\times$10$^{-4}$ & $-1.4\times$10$^{-3}$ \\

\hline
\end{tabular}
\end{table*}

For the Joule-heating-induced AC electrical response, an order of magnitude estimate of $\Im \Delta \rho^{1\omega}(j_0)$ can be obtained by referring to $\Re \Delta \rho^{1\omega}(j_0)$ and $\omega_c$ as follows: 
\begin{align}
\label{Im/Re}
\frac{\Im \Delta \rho^{1\omega}(\omega, j_0, T)}{\Re \Delta \rho^{1\omega}(\omega, j_0, T)} = - \frac{\chi''(2\omega, T)}{2\chi_0(T) + \chi'(2\omega, T)}.
\end{align}
Note that within the present model, which considers the power-linear term and disregards higher-order terms, the ratio of $\Im \Delta \rho^{1\omega}(\omega, j_0, T)$ to $\Re \Delta \rho^{1\omega}(\omega, j_0, T)$ does not depend on $j_0$. Using Eq.~(\ref{Debye}) with $\alpha =0$ for simplicity, we obtain:
\begin{align}
\label{Im1w/w_order}
\frac{\Im \Delta \rho^{1\omega}(\omega, j_0, T)}{\omega} \approx - \frac{1}{3} \frac{\Re \Delta \rho^{1\omega} (j_0, T)}{\omega_c} \hspace{0.5cm} \rm for \hspace{0.2cm} \omega < \omega_c,
\end{align}
where $\Re \Delta \rho^{1\omega}(j_0, T)$ represents the low-frequency limit value. By referring to the $\Re \Delta \rho^{1\omega}(j_0, T)$ and $\omega_c$ reported for Gd$_3$Ru$_4$Al$_{12}$ \cite{YokouchiNature}, YMn$_6$Sn$_{6}$ \cite{KitaoriPNAS}, and FeSn$_2$ \cite{YokouchiArxiv}, we calculate the $\Im \Delta\rho^{1\omega}(\omega, j_0, T)/\omega$ at low $\omega$ ($< \omega_c$) for each system. The results are summarized in Table I. The calculated values are in agreement with the reported data for all three systems. This correspondence on the order of magnitude indicates that the origin of the nonlinear $\Im \Delta\rho^{1\omega}(j_0)$ observed in the experiments \cite{YokouchiNature, KitaoriPNAS, YokouchiArxiv} lies in the delay of the nonlinear $\Re \Delta\rho^{1\omega}(j_0)$ with a cutoff frequency $\omega_c$. In other words, the above analysis suggests that the reported $\Im \Delta\rho^{1\omega}(j_0)$ does not represent the EEF-derived emergent inductance, and therefore, the $\Im \Delta\rho^{1\omega}(j_0)$ reported previously can be markedly larger than that expected based on the EEF mechanism. This is the answer to question I. Incidentally, Figs.~\ref{Gd3Ru4Al12}, \ref{Scale} and \ref{Hdep} indicate $\Re \Delta \rho^{1\omega}(j_0) \propto \Im \Delta \rho^{1\omega}(j_0) \propto -{\rm d}\rho_0/{\rm d} T$, and thus, the $\Re\Delta\rho^{1\omega}(j_0)$ is also naturally explained by considering Joule heating [Eq.~(\ref{Re1w})].

\section{V. Discussion}

In general, dissipative and nondissipative mechanisms can coexist in impedance. Phenomenologically, the nonlinear $\Delta \rho(j_0)$ is thus likely described by the sum of the two mechanisms: $\Delta \rho(j_0) = \Delta \rho_{\rm diss}(j_0) + \Delta \rho_{\rm nondiss}(j_0)$, where $\Delta \rho_{\rm diss}$ and $\Delta \rho_{\rm nondiss}$ represent the nonlinear impedances due to dissipative mechanisms and nondissipative mechanisms, respectively. Note that by definition, at low $\omega$ close to DC, $\Re \Delta \rho_{\rm diss} \gg \Im \Delta \rho_{\rm diss}$, and $\Im \Delta \rho_{\rm nondiss} \gg \Re \Delta \rho_{\rm nondiss}$. As mentioned in Sec.~II~C, the inductor mechanism due to the EEF caused by a pinned magnetic texture assumes $\Im \Delta \rho^{1\omega} \gg \Re \Delta \rho^{1\omega}$ at low $\omega$ and therefore belongs to $\Delta \rho_{\rm nondiss}$, whereas the Joule-heating-induced AC electrical response satisfies $\Re \Delta \rho^{1\omega} \gg \Im \Delta \rho^{1\omega}$ at low $\omega$ and therefore belongs to $\Delta \rho_{\rm diss}$. As shown in Sec.~IV, we have discovered that regardless of whether the material under consideration exhibits a magnetic order (as in \cite{YokouchiNature, KitaoriPNAS, YokouchiArxiv}) or not (as in this study), the observed nonlinear $\Delta \rho(j_0)$ is categorized into $\Delta \rho_{\rm diss}$ and has the characteristics of the Joule-heating-induced AC electrical response [Eqs.~(\ref{Re1w})--(\ref{Im3w})]. This finding implies that unless the time-varying Joule heating is negligibly small, any nondissipative signals that may coexist are easily masked by the Joule-heating-induced AC electrical response. Note that the nonlinear $\Im \Delta \rho^{1\omega}/\omega$ becomes pronounced when large AC currents are used at low frequencies for impedance measurements, as indicated by Eq.~(\ref{Im1w}). Therefore, when exploring nondissipative signals in conducting materials, it is important to avoid a nonlinear low-frequency regime and to suppress the time-varying Joule heating as much as possible. Since an order of magnitude estimate of the Joule-heating-induced $\Im \Delta \rho^{1\omega}(j_0)$ can be obtained if the amount of the temperature increase due to the Joule heating and the thermal-response time are determined from experimental results, it would also be important to double-check that the nondissipative signals thus obtained are greater than this estimated value.

\section{VI. Conclusion}
To clarify fundamental questions that have remained unanswered in the previous experiments on emergent inductors, we have considered the impact of time-varying Joule heating on the AC electrical responses. From a theoretical point of view, several key characteristics of the Joule-heating-induced AC electrical response within a power-linear regime have been clarified. To further examine the Joule-heating-induced AC electrical response, we have performed experiments on two materials that exhibit no magnetic order, CuIr$_2$S$_4$ and 1$T$'-MoTe$_2$, and verified the characteristics of the Joule heating model. We have reconsidered the nonlinear emergent inductance previously reported and determined that the temperature and magnetic field dependences, cutoff frequency, and magnitude of $\Im \rho^{1\omega}/\omega$ can be naturally explained in terms of the Joule-heating-induced AC electrical response. 

The Joule-heating-induced AC electrical response inevitably yields finite $\Im \rho^{1\omega}/\omega$ and $\Im \rho^{3\omega}/\omega$, unless the Joule heating is negligible and the measurement frequency is far greater than the inverse of the thermal response time. Even temperature oscillations as small as 0.1 K may cause a $\Im \Delta \rho^{1\omega}/\omega$ of considerable magnitude, which is much larger than that expected from the linear-response EEF. In previous experiments on emergent inductors, the nonlinear $\Im \rho^{1\omega}/\omega$ below the cutoff frequency was discussed, and the nonlinearity more pronouncedly occurred in real impedance than in imaginary impedance. Our results thus suggest that the reported data regarding emergent inductors need to be reconsidered by taking into account the impact of the Joule-heating-induced AC electrical response.

\textit{Note added in proof.} During the review process, our manuscript was commented by Yokouchi \textit{et al.}~from Tokura group \cite{Condmat}. After careful consideration of their comment and new data, we found no need to change our conclusion. Our response to the comment is provided in Ref.~\cite{Reply}.


\subsection*{A\lowercase{CKNOWLEDGMENTS}}
F.K., S.F. and W.K. thank T. Yokouchi and A. Kitaori for providing us with the raw data from the literature \cite{YokouchiNature, KitaoriPNAS}. F.K., S.F. and W.K. thank T. Yokouchi and S. Maekawa for fruitful discussions. We thank Y. Tokumura for the synthesis of CuIr$_2$S$_4$. This work was partially supported by JSPS KAKENHI (Grants No.~21H04442 and No.~23K03291), JST CREST (Grants No.~JPMJCR1874 and No.~JPMJCR20T1).

\section{APPENDIX}

\subsection{1. Methods}
\begin{figure*}
\includegraphics{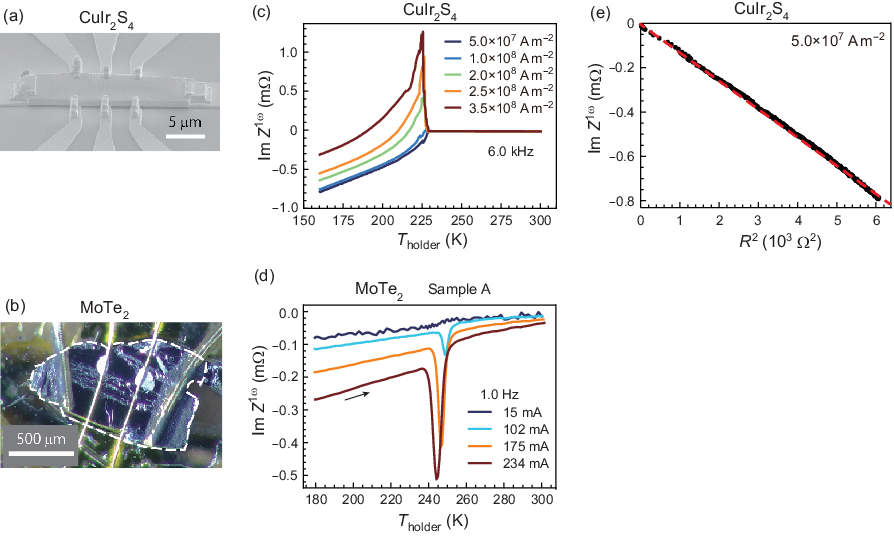}
\caption{\label{Photo} (a) Scanning electron microscopy image of the microfabricated CuIr$_2$S$_4$ device used in this study. (b) Photograph of the bulk MoTe$_2$ used in this study. (c) Raw data of the $\Im \rho^{1\omega}$--$T$ profile of the microfabricated CuIr$_2$S$_4$. (d) Raw data of the $\Im Z^{1\omega}$--$T$ profile of the microfabricated MoTe$_2$. (e) $R^2$ vs $\Im Z^{1\omega}$ plot for the microfabricated CuIr$_2$S$_4$ under a low AC current density of 5.0$\times10^7$ A~m$^{-2}$.}
\end{figure*}

The images of the microfabricated CuIr$_2$S$_4$ device and the bulk MoTe$_2$ are shown in Figs.~\ref{Photo}(a) and (b), respectively. In the experiments on MoTe$_2$, we used carbon paste for the current electrodes to facilitate the Joule heating; the resistivity of MoTe$_2$ is too low to achieve a Joule heating in bulk with use of $\sim$100 mA for the case of bulk crystal.

Figures \ref{Photo}(c) and (d) show the raw data of the $\Im \rho^{1\omega}$--$T$ profile of the microfabricated CuIr$_2$S$_4$ device and the $\Im Z^{1\omega}$--$T$ profile of the bulk MoTe$_2$ crystal, respectively. The linear-response background signal is nonmonotonically $T$ dependent in the microfabricated CuIr$_2$S$_4$ device, whereas it is relatively small and weakly $T$ dependent in the bulk MoTe$_2$ crystal. The linear-response background in the CuIr$_2$S$_4$ device is likely due to the impact of the stray capacitance $C$, which generates a linear-response background of $-i\omega R_0^2C$ \cite{Furuta1}, where $R_0$ is the linear-response DC resistance. The $\Im Z^{1\omega}$ is proportional to $R_0^2$ in the present temperature range [Fig.~\ref{Photo}(e)], indicating that the $(-R_0^2C)$-type background dominates the imaginary part of the linear-response signal in the microfabricated sample. Since the nonlinearly enhanced electrical response is the main focus of this study, we present $\Im \Delta Z^{1\omega}$, which represents a nonlinear change from the linear-response value.

\subsection{2. Nonlinear AC electrical response of MoTe$_2$ bulk crystal}

MoTe$_2$ shows a distinct change in ${\rm d}R_0/{\rm d}T$ through a first-order structural phase transition at $T_{\rm c} \approx 250$ K \cite{ZhouNatCommun}, and thus, this feature is helpful to study the correlation between $\Im Z^{1\omega}$ and $-{\rm d}R_0/{\rm d}T$.

\begin{figure*}
\includegraphics{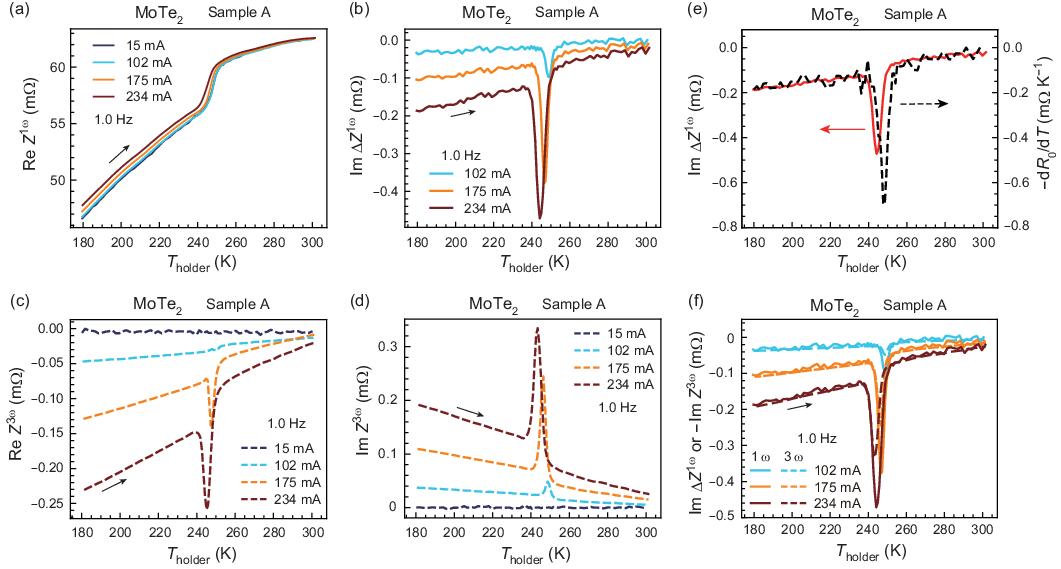}
\caption{\label{MoTe2} Temperature dependence of AC electrical response of MoTe$_2$ measured at various current amplitudes. (a) $\Re Z^{1\omega}$, (b) $\Im Z^{1\omega}$, (c) $\Re Z^{3\omega}$, and (d) $\Im Z^{3\omega}$. The data were recorded in the heating process. (e) Comparison between $\Im \Delta Z^{1\omega}$ and $-{\rm d}R_0/{\rm d}T$. (f) Comparison between $\Im \Delta Z^{1\omega}$ and $-\Im Z^{3\omega}$. }
\end{figure*}

Figures \ref{MoTe2}(a), (b), (c), and (d) display the temperature dependences of $\Re Z^{1\omega}$, $\Im \Delta Z^{1\omega}$, $\Re Z^{3\omega}$, and $\Im Z^{3\omega}$ at $\omega$/2$\pi = 1$ Hz, respectively, measured using various AC current amplitudes. In the $\Re Z^{1\omega}$--$T$ profile [Fig.~\ref{MoTe2}(a)], the apparent transition temperature clearly decreases as the current increases, indicating that the sample temperature increases from the sample-holder temperature, $T_{\rm holder}$, by Joule heating. Figs.~\ref{MoTe2}(b)--(f) show characteristic features consistent with the Joule heating model. First, $\Im \Delta Z^{1\omega}$, $\Re Z^{3\omega}$ and $\Im Z^{3\omega}$ nonlinearly emerge as the AC current increases [Figs.~\ref{MoTe2}(b)--(d)]. Second, the $\Im \Delta Z^{1\omega}$--$T$ profile qualitatively agrees with the $(-{\rm d}R_0/{\rm d}T)$--$T$ profile [Fig.~\ref{MoTe2}(e)], which is consistent with the results expected when Joule heating occurs mainly at the contacts. Third, the relation of $\Im \Delta Z^{1\omega} = -\Im Z^{3\omega}$ is well satisfied, with the exception of the transition region at 245 K [Fig.~\ref{MoTe2}(f)]. The breakdown of $\Im \Delta Z^{1\omega} = -\Im Z^{3\omega}$ at 245 K appears reasonable considering that Eqs.~(\ref{Re1w})--(\ref{Im3w}) do not generally describe the data at and near a phase transition well.

\begin{figure*}
\includegraphics{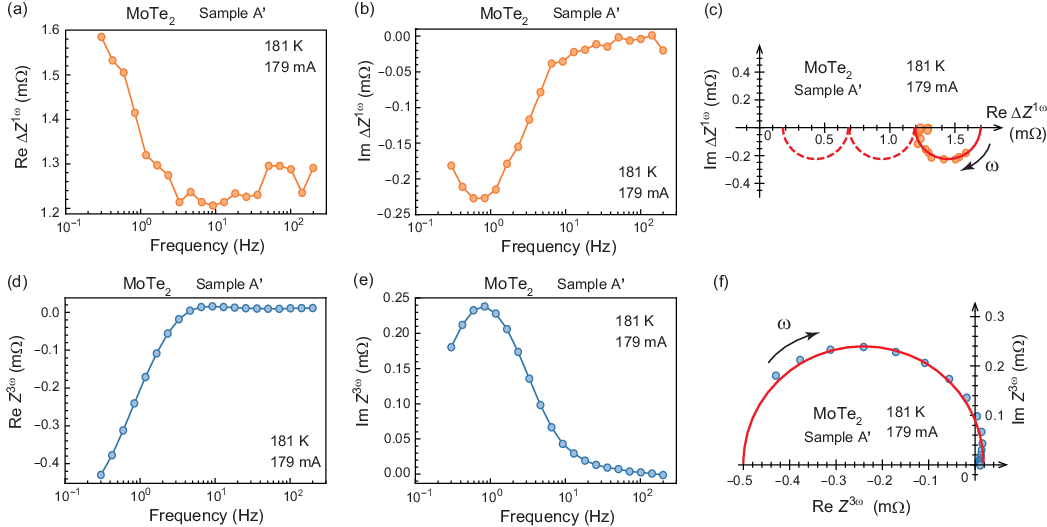}
\caption{\label{MoTe2_freq} Frequency dependence of the nonlinear AC electrical response of MoTe$_2$ at 181 K. (a) $\Re \Delta Z^{1\omega}$, (b) $\Im \Delta Z^{1\omega}$, and (c) Cole-Cole representation of $\Delta Z^{1\omega}$. (d) $\Re Z^{3\omega}$, (e) $\Im Z^{3\omega}$, and (f) Cole-Cole representation of $Z^{3\omega}$. The data were recorded using an AC current of $j_0=179$ mA. As the experiments progressed, the contact resistance changed from the state shown in Fig.~\ref{MoTe2}, and the sample is therefore labelled Sample A' instead of A. The geometries of Sample A and Sample A' are exactly the same.}
\end{figure*}

The frequency dependences and Cole-Cole representations of $\Delta Z^{1\omega}$ and $Z^{3\omega}$ at 180 K are shown in Figs.~\ref{MoTe2_freq}(a)--(f). They are also consistent with the predictions of the Joule heating model for the case of ${\rm d}R_0/{\rm d}T > 0$, although $\Delta Z^{1\omega}$ appears to be more susceptible to the linear-response background than $Z^{3\omega}$ is. In the Cole-Cole representations of $\Delta Z^{1\omega}$ and $Z^{3\omega}$, the lengths of the arc strings are approximately the same (0.5 m$\Omega$). The cutoff frequency of the nonlinear AC electrical response is estimated as $\approx$1 Hz. This frequency is typical for the thermal response in a bulk crystal, as often reported in AC-temperature calorimetry experiments \cite{ACthermal1}.



\clearpage

%

\renewcommand{\figurename}{Fig.~S}
\setcounter{figure}{0}
\bibliographystyle{apsrev}

\begin{thebibliography}{99}
\bibitem{Note} Note that this comparison is not necessarily an ideal verification of the Joule-heating model up to the power-linear order [Eq.~(\ref{Im1w})] because Fig.~\ref{Gd3Ru4Al12}(a) indicates that ${\partial}\rho(j_0)/{\partial}T$ at 5 K pronouncedly depends on $j_0$, which is beyond the scope of assumption $\#$4 of the model.
\end{thebibliography}
\def\Vec#1{\mbox{\boldmath $#1$}}


\section{Supplemental Material}



\begin{figure*}[b]
\includegraphics{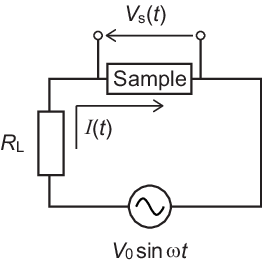}
\caption{Typical experimental circuit used for measurement of the complex impedance.}
\end{figure*}

In Supplemental Material, we show that the negative nonlinear reactance reported in previous experimental studies \cite{YokouchiNature, KitaoriPNAS, YokouchiArxiv} is not a manifestation of negative nonlinear inductance. In previous experimental reports, the impedance at the fundamental and higher hamonic frequencies was measured for the input frequency $\omega/(2\pi)$ using the lock-in technique. The typical experimental setup is displayed in Fig.~S1: a magnetic material, a load resistor with resistance, $R_{\rm L}$, and an AC voltage output of the lock-in amplifier, $V_0 \sin \omega t$, are connected in series. For simplicity, we consider a sample with a two-probe configuration and do not explicitly consider the contact resistance. The current flowing through the circuit, $\mathcal{I}^{1\omega}(\omega)$ [the $1\omega$ Fourier component of $I(t)$], was determined from the voltage drop at the load resistor, and the voltage drop at the sample, $\mathcal{V}^{1\omega}_{\rm s} (\omega)$ [the $1\omega$ Fourier component of $V_{\rm s}(t)$], was measured. Then, they defined the complex impedance $Z^{1\omega}(\omega,\mathcal{I})$ at $\mathcal{I}^{1\omega}(\omega)$ by $\mathcal{V}^{1\omega}_{\rm s} (\omega)/\mathcal{I}^{1\omega}(\omega)$. Note that this definition was also used for a large $\mathcal{I}^{1\omega}(\omega)$ such that the proportionality between $\mathcal{V}^{1\omega}_{\rm s}(\omega)$ and $\mathcal{I}^{1\omega}(\omega)$ is no longer valid. From a theoretical perspective, they defined an emergent inductor as an element showing the following voltage drop:
\begin{equation}
V_{\rm s}(t) = R_{\rm s}I(t)+L_0 \Bigl( 1+{\rm A}I(t)^2 +{\rm B}I(t)^4 + \cdots \Bigr) \frac{{\rm d}I}{{\rm d}t}, \tag{S1}
\end{equation}
where $R_{\rm s}$ is the resistance of the sample, $L_0$ denotes the inductance in the linear response, and A and B are coefficients representing the nonlinearity related to the inductive response. From an energetic perspective, Eq.~(S1) corresponds to the fact that under current $I$, the inductive element with a finite resistance can store the following energy:
\begin{equation}
E_{\rm L} \bigl( I(t) \bigr) = L_0 \left( \frac{1}{2}I(t)^2 + \frac{1}{4}{\rm A}I(t)^4 + \frac{1}{6}{\rm B}I(t)^6 +\cdots \right). \tag{S2}
\end{equation}
Note that the inductive term in Eq.~(S1), multiplied by $I$, is equal to ${\rm d}E_{\rm L}/{\rm d}t$, indicating the relationship between the inductive electric response and the energy stored in the inductor. Given Eq.~(S1), the circuit equation of the measurement system is given as:
\begin{align}
V_0 \sin \omega t &= (R_{\rm L}+R_{\rm s})I(t) +L_0 \Bigl( 1+{\rm A}I(t)^2 +{\rm B}I(t)^4 + \cdots \Bigr) \frac{{\rm d}I}{{\rm d}t} \nonumber \\
&= RI(t) + L_0 \Bigl( 1+{\rm A}I(t)^2 +{\rm B}I(t)^4 + \cdots \Bigr) \frac{{\rm d}I}{{\rm d}t}, \tag{S3}
\end{align}
where $V_0$ is the output voltage amplitude of the lock-in amplifier and $R$ is the sum of the sample and load resistances. Experimentally, they observed $\Im [\mathcal{V}^{1\omega}_{\rm s} (\omega)/\mathcal{I}^{1\omega} (\omega)] \approx -| \eta | \omega$ at low frequencies, especially when the flowing current is large under a large $V_0$. This observation was interpreted as indicating that the coefficient of ${\rm d}I/{\rm d}t$ in Eq.~(S3) became negative under large currents. However, it has not been discussed whether Eq.~(S3) truly shows $\Im [\mathcal{V}^{1\omega}_{\rm s} (\omega)/\mathcal{I}^{1\omega} (\omega)] \approx -| \eta | \omega$ when the coefficient of ${\rm d}I/{\rm d}t$ is negative at a large $I$. Therefore, the correspondence between the experimental observations and Eqs.~(S1)--(S3) remains unclear. In fact, as shown below, the solution of Eq.~(S3) has instability towards self-sustained oscillations when the coefficient of ${\rm d}I/{\rm d}t$ becomes negative at a large $I$, and it does not show an electric response such that $\Im [\mathcal{V}^{1\omega}_{\rm s} (\omega)/\mathcal{I}^{1\omega} (\omega)] \approx -| \eta | \omega$; i.e., the experimental observations are not described by Eqs.~(S1)--(S3). In the following, we numerically examine the nature of the nonlinear negative inductance represented by Eqs.~(S1)--(S3).

Equation (S3) is complicated because the coefficient of ${\rm d}I/{\rm d}t$ is time dependent and may change the sign during the time evolution. To gain insights into Eq.~(S3), it would be instructive to begin with a simpler case. We consider a different form of circuit equation, which is nonlinear with respect to the input voltage amplitude, $V_0$, instead of the current:
\begin{align}
V_0 \sin \omega t &= RI(t) +L_0 ( 1+{\rm A_V}V_0^2 +{\rm B_V}V_0^4 + \cdots ) \frac{{\rm d}I}{{\rm d}t} \nonumber \\
&= RI(t) + L_{\rm V}(V_0)\frac{{\rm d}I}{{\rm d}t}. \tag{S4} 
\end{align}
Note that in Eq.~(S4), the coefficient of ${\rm d}I/{\rm d}t$, $L_{\rm V}(V_0)$, is constant during the time evolution of the system, and thus, the profile of $I(t)$ can be easily deduced from the knowledge of dynamic systems \cite{Textbook}. The numerical results are displayed in Figs.~S2(a) and (b). For clarity, here, we chose a simplified parameter set: $V_0 = 1, \omega = 1, R = 10, I(0) = 0$, and $L_{\rm V}(V_0) = +5$ [Fig.~S2(a)] or $-5$ [Fig.~S2(b)]. Note that the qualitative behavior of the solution depends only on whether the coefficient of ${\rm d}I/{\rm d}t$ is positive or negative, and we chose parameters such that the characteristics of $I(t)$ can be clearly observed. Figure S2(a) shows the result for the case of $L_{\rm V}(V_0) = +5$. When the coefficient of ${\rm d}I/{\rm d}t$ is positive, the focus, $I_0(t)=(V_0 \sin \omega t)/R$ (denoted by the dotted line), is a stable focus (but time dependent due to the AC driving force, $V_0 \sin \omega t$); thus, $I(t)$ tracks the time-varying focus, with a finite delay. Figure S2(b) shows the case of $L_{\rm V}(V_0) = -5$. When the coefficient is negative, the focus is unstable, and thus, $I(t)$ tends to separate from it over time. As a result, once $I(t)$ deviates from the focus by an infinitesimal amount, the deviation is unlimitedly amplified and diverges. Thus, a constant negative inductance makes the system unstable. This divergence instability is a natural consequence of the fact that the energy of the inductor, $\frac{1}{2} L_V (V_0) I^2$, diverges to negative infinity by increasing $|I|$ under constant negative inductance $L_V(V_0)$.

\begin{figure*}
\includegraphics{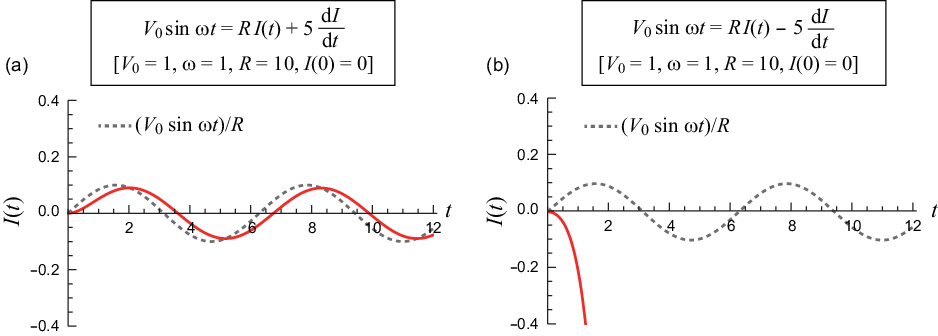}
\caption{Time evolution of the current in the differential equation, Eq.~(S4), for the case of positive inductance, $L_{\rm v}(V_0) = +5$ (a) and negative inductance, $L_{\rm v}(V_0) = -5$ (b).
}
\end{figure*}

\begin{figure}
\includegraphics{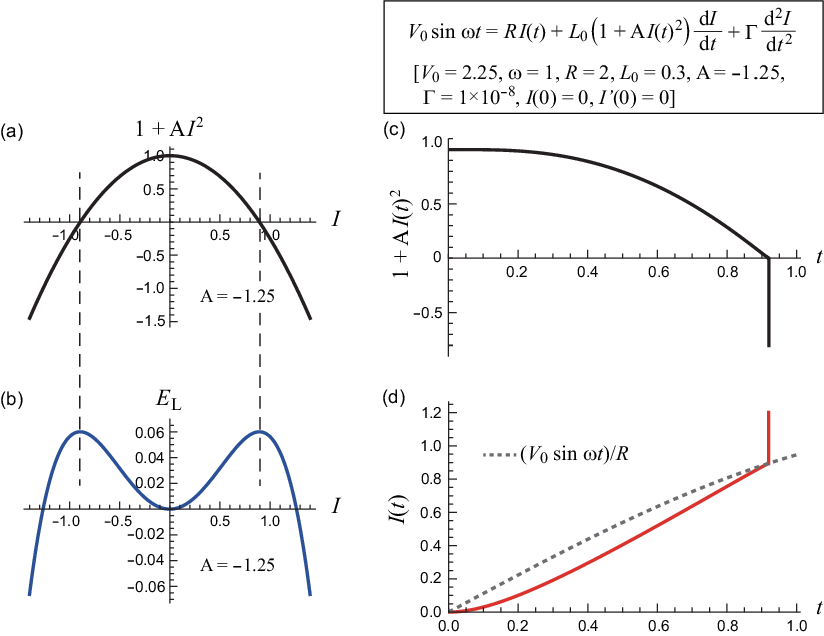}
\caption{Time evolution of the current in the differential equation, Eq.~(S5) with ${\rm B} = 0$. (a, b) The coefficient of ${\rm d}I/{\rm d}t$ (divided by $L_0$) (a) and the energy stored in the inductor (b) as a function of current. (c, d) Time evolutions of the coefficient of ${\rm d}I/{\rm d}t$ (c) and current (d).}
\end{figure}

We next consider the main issue, Eq.~(S3), which is a more nontrivial circuit equation in the sense that the coefficient of ${\rm d}I/{\rm d}t$ varies during the time evolution. We set positive $L_0$ $(= 0.3)$ so that the system is stable in the linear response. By choosing appropriate A and B, Eq.~(S3) represents a nontrivial situation such that the coefficient of ${\rm d}I/{\rm d}t$ changes its sign from positive to negative when the current becomes sufficiently large. Note that Eq.~(S3) can be rewritten as ${\rm d}I/{\rm d}t = (V_0 \sin \omega t - RI)/[L_0 (1+{\rm A}I^2+{\rm B}I^4)]$, and thus, it shows singularity the moment $L_0 (1+{\rm A}I^2+{\rm B}I^4)$ reaches zero. To numerically solve Eq.~(S3), we therefore introduce ${\rm d}^2 I/{\rm d}t^2$ with an infinitesimally small positive coefficient, $\Gamma$ ($=10^{-8}$ in the present numerical calculation), which is just for the sake of the stability of the numerical tracking of the solution and is not important in the following discussion. Namely, the differential equation actually computed in this Supplemental Material is:
\begin{align}
V_0 \sin \omega t = RI(t) + L_0 \Bigl( 1+{\rm A}I(t)^2 +{\rm B}I(t)^4 \Bigr) \frac{{\rm d}I}{{\rm d}t} + \Gamma \frac{{\rm d}^2 I}{{\rm d}t^2}. \tag{S5}
\end{align}
When considering real systems, it is also natural to assume that $I(t)$ is always first-order differentiable (i.e., ${\rm d}I/{\rm d}t$ is always finite), and it therefore makes sense to consider such a second-order derivative term to ensure first-order differentiability.

First, we consider the case in which ${\rm A} = -1.25$ and ${\rm B} = 0$; thus, the coefficient of ${\rm d}I/{\rm d}t$ and $E_{\rm L}$ depend on $|I|$, as shown in Figs.~S3(a) and (b), respectively. The coefficient of ${\rm d}I/{\rm d}t$ and ${\rm d} E_{\rm L} /{\rm d}I$ reach zero at $I \approx \pm 0.9$. Figures S3(c) and (d) show the numerical results in the nonlinear regime such that the coefficient of ${\rm d}I/{\rm d}t$ reaches zero: the parameter set used is: $V_0 = 2.25, \omega = 1, R = 2, L_0 = 0.3, \Gamma = 1\times10^{-8}, I(0) = 0$, and $I'(0) = 0$. As the current increases with time, the coefficient of ${\rm d}I/{\rm d}t$ decreases and eventually reaches zero at $t \approx 0.92$ [Fig.~S3(c)]. Concomitantly, ${\rm d}I/{\rm d}t \approx (V_0 \sin \omega t-RI)/[L_0 (1+{\rm A}I^2+{\rm B}I^4 )]$ diverges approximately positively so that the current catches up with the time-varying focus, $I_0(t) =(V_0 \sin \omega t)/R$, from below and overshoots it [Fig.~S3(d)]. This upwards overshoot causes the coefficient of ${\rm d}I/{\rm d}t$ to become negative [Fig.~S3(c)], which never reverts to a positive value. Thus, the time-varying focus is unstable for $t > 0.92$, and accordingly, the current continues to separate from it and eventually diverges, similar to the case of Fig.~S2(b). Thus, the system is unstable for a large $V_0$. This behavior reflects the fact that in the absence of the $I^4$ term in Eq.~(S5), the inductor energy can be unlimitedly decreased by increasing $|I|$ when the current exceeds the threshold value, $|I| \approx 0.9$ [Fig.~S3(b)].

\begin{figure}[b]
\includegraphics{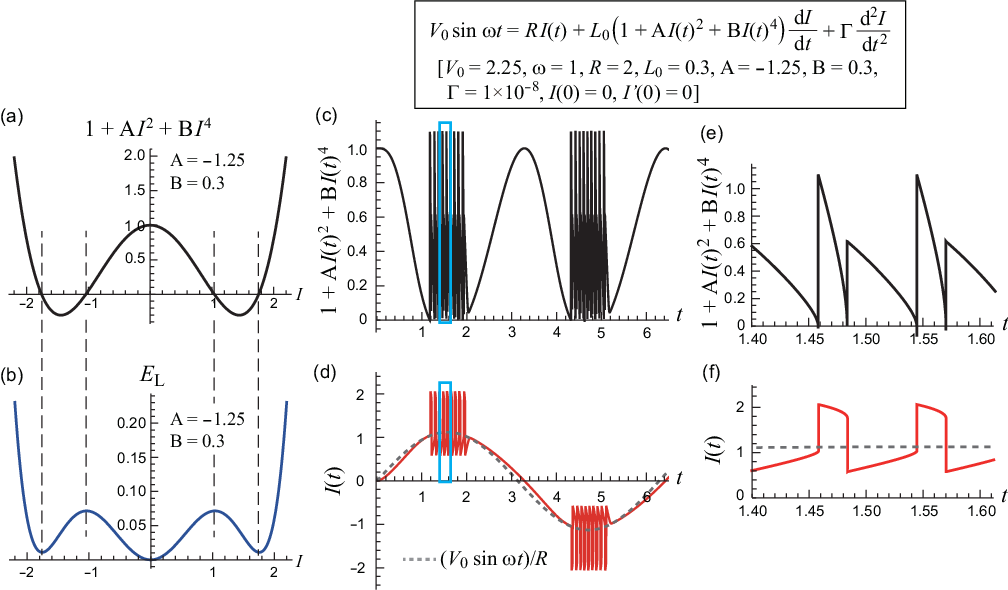}
\caption{Time evolution of the current in the differential equation, Eq.~(S5) with positive B. (a, b) The coefficient of ${\rm d}I/{\rm d}t$ (divided by $L_0$) (a) and the energy stored in the inductor (b) as a function of current. (c, d) Time evolutions of the coefficient of ${\rm d}I/{\rm d}t$ (c) and current (d). (e, f) Enlarged view of the squared area in panel (c) and (d), respectively.
}
\end{figure}

Next, we consider the case in which ${\rm A} = -1.25$ and ${\rm B} = 0.3$; i.e., the $I^4$ term with a positive coefficient is present in Eq.~(S5). Thus, as $|I|$ increases, the coefficient of ${\rm d}I/{\rm d}t$ first changes from positive to negative and then back to positive again [Fig.~S4(a)]; furthermore, for this parameter choice, the inductor energy is always positive for arbitrarily chosen $I$ [Fig.~S4(b)]. These features imply that the simultaneous divergence of $I$ and the coefficient of ${\rm d}I/{\rm d}t$, as observed in Figs.~S3(c) and (d), does not occur. Figures S4(c) and (d) show the numerical results in the nonlinear regime such that the coefficient of ${\rm d}I/{\rm d}t$ reaches zero: the parameter set used is: $V_0 = 2.25, \omega = 1, R = 2, L_0 = 0.3, \Gamma = 1\times10^{-8}, I(0) = 0$, and $I'(0) = 0$. When the coefficient of ${\rm d}I/{\rm d}t$ decreases from positive to zero, the system starts to oscillate [see also Figs.~S4(e) and (f), which are magnified views of Figs.~S4(c) and (d) during oscillation, respectively]. This oscillating behavior can be understood by considering the time evolution of Eq.~(S5) in steps as follows. Whenever the coefficient of ${\rm d}I/{\rm d}t$ reaches zero, the current tends to quickly catch up with the focus, $I_0 (t)=(V_0 \sin \omega t)/R$, overshoot it upwards from below (or downwards from above), and diverge while $1+{\rm A}I^2+{\rm B}I^4<0$ (i.e., $1<|I|< 1.75$); however, when the current falls outside this range, the coefficient of ${\rm d}I/{\rm d}t$ becomes positive again [Fig.~S4(a)]; thus, unlike in the case of ${\rm B} = 0$ [Figs.~S3(c) and (d)], the divergence of the current is halted, and the tracking towards the focus from above (or from below) restarts, resulting in decreasing (or increasing) current; during this tracking, the coefficient of ${\rm d}I/{\rm d}t$ decreases and reaches zero again. In this way, a jerky oscillation of the current is achieved around the time-varying focus [Figs.~S4(d) and (f)] when the current value results in a negative coefficient of ${\rm d}I/{\rm d}t$, exemplifying the unstable nature inherent to a negative coefficient of ${\rm d}I/{\rm d}t$.

\begin{figure*}
\includegraphics{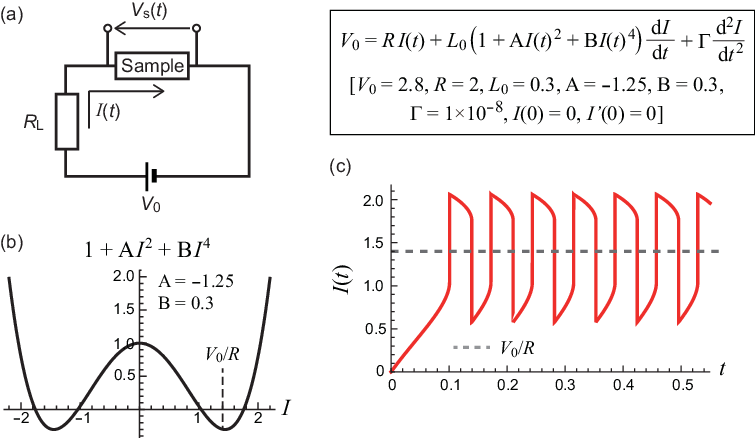}
\caption{Oscillator function of an element that shows a negative coefficient of ${\rm d}I/{\rm d}t$ under the application of a DC-voltage of appropriate magnitude. (a) Experimental circuit for the demonstration of the oscillator function. (b) The coefficient of ${\rm d}I/{\rm d}t$ (divided by $L_0$) as a function of current. The broken line indicates a condition of $V_0$ and $R$ that show the oscillator function. (c) Time evolution of the self-sustained current oscillation in the differential equation, Eq.~(S5) with positive B.
}
\end{figure*}

In summary, numerical examinations have demonstrated that the hallmark of a negative coefficient of ${\rm d}I/{\rm d}t$ is instability, such as divergence or spontaneous oscillation, and this phenomenon is never observed as an impedance with negative reactance. As would be best demonstrated by connecting a DC-voltage source of appropriate magnitude, an element that has a negative coefficient of ${\rm d}I/{\rm d}t$ in a certain current range will operate as an oscillator [Figs.~S5(a)--(c)]. Therefore, the experimental observation, $\Im [\mathcal{V}^{1\omega}_{\rm s} (\omega)/\mathcal{I}^{1\omega} (\omega)] \approx -| \eta | \omega$, in the nonlinear regime should be considered beyond the framework of Eqs.~(S1)--(S3). As discussed in the main text, we consider this issue in terms of time-varying Joule heating and its impact on the AC electrical response.



\end{document}